\documentclass[aps,prc,reprint,amsmath,amssymb,showpacs,preprintnumbers,superscriptaddress]{revtex4-1}
\usepackage{CJK}
\usepackage{graphicx}
\usepackage{dcolumn}
\usepackage{bm}
\usepackage{color}
\usepackage{hyperref}

\begin{document}

\begin{CJK*}{UTF8}{}

\title{Dynamics of rotation in chiral nuclei} 
\author{Z. X. Ren \CJKfamily{gbsn} (任政学) }
\affiliation{State Key Laboratory of Nuclear Physics and Technology, School of Physics, Peking University, Beijing 100871, China}

\author{P. W. Zhao \CJKfamily{gbsn}(赵鹏巍) }
\email{pwzhao@pku.edu.cn}
\affiliation{State Key Laboratory of Nuclear Physics and Technology, School of Physics, Peking University, Beijing 100871, China}

\author{J. Meng \CJKfamily{gbsn}(孟杰) }
\email{mengj@pku.edu.cn}
\affiliation{State Key Laboratory of Nuclear Physics and Technology, School of Physics, Peking University, Beijing 100871, China}

\begin{abstract}
 The dynamics of chiral nuclei is investigated for the first time with the time-dependent and tilted axis cranking covariant density functional theories on a three-dimensional space lattice in a microscopic and self-consistent way.
 The experimental energies of the two pairs of the chiral doublet bands in $^{135}$Nd are well reproduced without any adjustable parameters beyond the well-defined density functional.
 A novel mechanism, i.e., chiral precession, is revealed from the microscopic dynamics of the total angular momentum in the body-fixed frame, whose  harmonicity is associated with a transition from the planar into aplanar rotations with the increasing spin.
 This provides a fully microscopic and dynamical view to understand the chiral excitations in nuclei.
\end{abstract}


\maketitle

\end{CJK*}

Chirality is a well-known phenomenon in many fields, such as chemistry, biology, molecular and particle physics.
In nuclear physics, chirality was originally suggested by Frauendorf and Meng in 1997~\cite{FRAUENDORF1997Chiral}.
It represents a novel feature of triaxial nuclei rotating around an axis which lies outside the three planes spanned by the principal axes of the triaxial ellipsoidal density distribution, i.e., \emph{aplanar rotation}.
The total angular momentum consists of three components along the short, intermediate, and long principal axes of a triaxial nucleus in the body-fixed frame, and they form either a left- or right-handed system.
The two chiral systems are related by the chiral operator $TR(\pi)$ that combines time reversal $T$ and spatial rotation by $\pi$.
The broken chiral symmetry in the body-fixed frame should be restored in the laboratory frame.
This leads to the so-called chiral doublet bands, which consist of a pair of nearly-degenerate $\Delta I = 1$ rotational bands~\cite{FRAUENDORF1997Chiral,Frauendorf2001RMP}.
The evidences of chiral doublet bands have been reported experimentally in the $A\approx80$~\cite{Wang2011Br80, Liu2016Br78}, 100~\cite{Vaman2004Rh104, Joshi2004Rh106, Tonev2014Rh102, Lieder2014Ag106, Rather2014Ag106, Kuti2014Rh103}, 130~\cite{Starosta2001Chiral, Zhu2003Nd135, Tonev2006Pr134, Petrache2006Pr134_Pm136, Grodner2006Cs128, Tonev2007Pr134, Mukhopadhyay2007Nd135, Ayangeakaa2013Ce133, Petrache2018Nd136, Lv2019PRCNd135}, and 190~\cite{Balabanski2004Ir188, Lawrie2008Tl198} mass regions; see also data tables~\cite{XIONG2019Chiral_Data}.

The manifestation of chirality in realistic nuclei is usually more complicated~\cite{Meng2010chiral, Meng2016CDFT_chiral_MR}.
In most cases, the chiral doublet bands are more separated in energy at low spins than at high spins.
At low spins, the energy difference between the states with the same spin in the chiral partners is caused by rapid vibration between the left- and right-handed configurations, i.e., \emph{chiral vibration}.
When the spin grows above a certain critical value, the energy splittings between the partner bands decrease, and the left-right mode changes from soft chiral vibration to tunneling between well-established chiral configurations, i.e., \emph{static chirality}.

On the theoretical side, many phenomenological approaches have been employed to understand such an intriguing phenomenon, such as the particle rotor model~\cite{FRAUENDORF1997Chiral, Koike2004ChiralPRM, Peng2003ChiralPRM, Zhang2007ChiralPRM, Qi2009ChiralPRM, Chen2018ChiralPRM}, the generalized coherent state model~\cite{Raduta2016Chiral}, and the interacting boson fermion-fermion model~\cite{Tonev2006Pr134, Tonev2007Pr134, Brant2008ChiralIBM}.
However, they are based on several assumptions, the validity of which may not be assured, and are fitted to the data in one way or another.

A microscopic treatment is thus important for a better understanding of nuclear chirality.
The tilted axis cranking (TAC) approach allows one to study nuclear chirality based on a microscopic mean field.
Although it well describes the energies and the intraband transition rates of the lower one of the chiral doublet bands, it gives either one achiral solution or two degenerate chiral ones~\cite{Dimitrov2000Chirality}, so it cannot describe the left-right excitation mode.
This is a well-known deficiency of the mean-field solution when it spontaneously breaks a symmetry~\cite{Frauendorf2001RMP}.
Consequently, the random-phase approximation method~\cite{Mukhopadhyay2007Nd135, Almehed2011ChiralVibration}, the collective Hamiltonian method~\cite{Chen20131DCH, Chen20162DCH}, and the projected shell model~\cite{Bhat2012Chiral, Chen2017Chiral, Chen2018Chiral, Wang2019Chiral} have been developed to describe the chiral excitations.
However, these approaches are based on single-particle potentials combined with either the schematic single-$j$ Hamiltonian~\cite{Chen20162DCH} or the pairing plus quadrupole-quadrupole model~\cite{Frauendorf2000PPQQ}.
Self-consistent methods based on more realistic two-body interactions are required for a more fundamental investigation, including all important effects such as core polarization and nuclear currents~\cite{Zhao2011PRL_AMR, Meng2013FT_TAC}.

Moreover, all the previous studies on nuclear chirality are based on static approaches, so the dynamics of the chiral excitations is still missing.

The nuclear density functional theory (DFT) starts from a universal energy density functional and can achieve a self-consistent description for almost all nuclei.
Covariant density functional theory (CDFT) further exploits the Lorentz symmetry, so it includes the complicated interplay between the large Lorentz scalar and vector self-energies~\cite{Volum16,Ren2020SelfSRG}.
It also allows the self-consistent treatment of the spin degrees of freedom and the nuclear currents induced by the spatial parts of the vector self-energies, which play an essential role in rotating nuclei; see Refs.~\cite{Meng2013FT_TAC, meng2016relativistic} for details.
Both relativistic and nonrelativistic DFTs have been extended with the TAC method to investigate nuclear chirality~\cite{Olbratowski2004crititcal, Zhao2017ChiralRotation}.
In particular, the recent TAC-CDFT has been widely used to investigate the chirality in nuclei $^{106}\mathrm{Rh}$~\cite{Zhao2017ChiralRotation}, $^{135}\mathrm{Nd}$~\cite{PENG2020Nd135}, $^{136}\mathrm{Nd}$~\cite{Petrache2018Nd136}, and $^{106}\mathrm{Ag}$~\cite{Zhao2019ChiralAg106}.
However, in all these works, only the lower band of the chiral doublet bands can be calculated.

The time-dependent DFT is a dynamical extension of DFT, and has been widely applied to various nuclear structure and reaction processes, see recent reviews~\cite{NakatsukasaRMP2016, SIMENEL2018TDHF_PPNP, STEVENSON2019PPNP} for details.
The solution of the time-dependent covariant density functional theory (TDCDFT) in three-dimensional lattice space is feasible only recently~\cite{Ren2020HeBeTDCDFT, Ren2020TDCDFT_O16+O16} thanks to the progress for solving  CDFT in the same space~\cite{tanimura20153d, REN2017Dirac3D, Ren2019C12LCS, REN2020Toroidal}.

\begin{figure}[!htbp]
  \centering
  \includegraphics[width=0.35\textwidth]{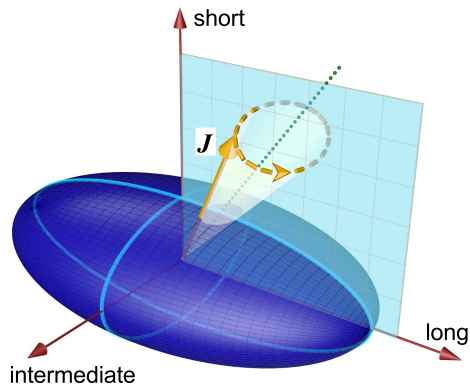}\\
  \caption{(Color online). A schematic picture for the ``chiral precession'' of a triaxial nucleus in the body-fixed frame,
  where the short, intermediate, and long axes are shown explicitly.
  The total angular momentum $\bm{J}$ of the nucleus is rotating about an axis (dotted line) in the body-fixed frame.
  }\label{fig1}
\end{figure}

In this Letter, for the first time, the dynamics of chiral nuclei is studied with the time-dependent and tilted axis cranking CDFT in a microscopic and self-consistent way.
By taking $^{135}$Nd as an example, the experimental energies of the two pairs of the chiral doublet bands are well reproduced without any adjustable parameters beyond the well-defined density functional.
In particular, a novel mechanism, \emph{chiral precession} (see Fig.~\ref{fig1}), is clearly revealed from the microscopic dynamics of the nuclear chiral excitations.

The starting point of CDFT is a universal energy density functional~\cite{RING1996PPNP, Vretenar2005PhysicsReport, meng2006PPNP, NIKSIC2011PPNP, meng2016relativistic}.
For nuclear chirality, the functional is transformed into a body-fixed frame rotating with a constant angular velocity vector $\bm{\omega}$, which is along an arbitrary direction in space, i.e., the three-dimensional TAC-CDFT~\cite{Zhao2017ChiralRotation}.
The corresponding Kohn-Sham equation for nucleons is a static Dirac equation,
\begin{equation}\label{eq_KS}
  \left[\bm{\alpha}\cdot(\hat{\bm{p}}-\bm{V})+V^0+\beta(m+S)-\bm{\omega}\cdot\hat{\bm{J}}\right]\psi_k(\bm{r}) = \varepsilon_k\psi_k(\bm{r}),
\end{equation}
where $\hat{\bm{J}}$ is the angular momentum, and the relativistic scalar $S(\bm{r})$ and vector $V^\mu(r)$ fields are connected in a self-consistent way to the nucleon densities and current distributions, which are obtained from the Dirac spinors $\psi_k(\bm{r})$.

In this work, the density functional PC-PK1~\cite{Zhao2010PC-PK1} is employed, and the calculations are free of additional parameters.
The Dirac equation \eqref{eq_KS} is solved in a cubic box with the length $24$ fm and the mesh size $0.8$ fm.
We focus on the chirality in the odd-$A$ nucleus $^{135}\mathrm{Nd}$, which was first observed in Ref.~\cite{Zhu2003Nd135} with the configuration consisting of two proton particles and one neutron hole in the $h_{11/2}$ shell, i.e., $\pi h_{11/2}^2\otimes\nu h_{11/2}^{-1}$.
Recently, a new pair of chiral doublet bands were reported with the configuration $\pi[h_{11/2}^1(gd)^1]\otimes\nu h_{11/2}^{-1}$~\cite{Lv2019PRCNd135}, which makes it a new candidate for multiple chirality~\cite{Meng2006MxD}.
In the following, we take the positive-parity configuration as an example to illustrate the analyses.

\begin{figure}[!htbp]
  \centering
  \includegraphics[width=0.45\textwidth]{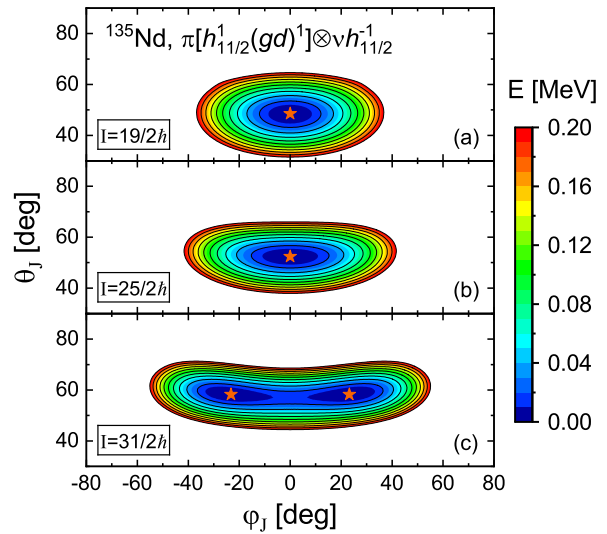}\\
  \caption{(Color online). Total energy surfaces for $^{135}$Nd with respect to the tilted angles $(\theta_J,\varphi_J)$ for the total angular momenta at $I=19/2\hbar$ (top), $25/2\hbar$ (middle), and $31/2\hbar$ (bottom) by the TAC-CDFT calculations based on the configuration $\pi[h_{11/2}^1(gd)^1]\otimes\nu h_{11/2}^{-1}$. The minima of the energy surfaces are denoted as stars.
  }\label{fig2}
\end{figure}

In Fig.~\ref{fig2}, the total energy surfaces are depicted in the plane of the tilted angles $\theta_J$ and $\varphi_J$ of the total angular momentum. Here, $\theta_J$ is the polar angle between the total angular momentum and the long axis, and $\varphi_J$ is the azimuth angle between the projection of the total angular momentum in the short-intermediate plane and the short axis.
At $I=19/2\hbar$, the total energy surface has a minimum at $\theta_J=48.5^\circ$ and $\varphi_J=0^\circ$.
This means that the total angular momentum is in the short-long plane, i.e., planar rotation.
At $I=25/2\hbar$, the location of the energy minimum varies only slightly in the $\theta$ direction, but the energy surface becomes much softer in the $\varphi$ direction.
This indicates that it would cost less to vibrate the angular momentum direction away from $\varphi_J=0^\circ$, i.e., the so-called chiral vibration, in comparison with the case at $I=19/2\hbar$.
At $I=31/2\hbar$, two degenerate minima locate in the energy surface at the same $\theta_J$ while opposite $\varphi_J$ values, which correspond to the left- and right-handed states, respectively.
Note that the TAC-CDFT gives either one planar solution or two degenerate aplanar ones, but it cannot describe the left-right excitation mode. Therefore, the TDCDFT calculations are further carried out.

In TDCDFT, the time-dependent Kohn-Sham equation reads~\cite{Ren2020HeBeTDCDFT, Ren2020TDCDFT_O16+O16},
\begin{equation}\label{eq_TDKS}
  i\partial_t\psi_k(\bm{r},t) = [\bm{\alpha\cdot}(\hat{\bm{p}}-\bm{V})+V^0+\beta(m+S)]\psi_k(\bm{r},t),
\end{equation}
where the time-dependent fields $S(\bm{r},t)$ and $V^\mu(\bm{r},t)$ have the same dependence on density and currents as in TAC-CDFT, so there are no additional parameters introduced.
In this work, we solve Eq.~\eqref{eq_TDKS} in the same lattice space as in the TAC-CDFT calculations, and the initial states are taken from the self-consistent solutions of the TAC-CDFT calculations.

Starting from the TAC-CDFT solutions, one can obtain the dynamical evolution of the system by means of the time-dependent wave functions $\psi_k(\bm{r},t)$.
Note that the solutions of TDCDFT are in the space-fixed frame, where the angular momentum is conserved during the time evolution, but the nuclear density distribution varies.
Since the nuclear deformation barely changes during the time evolution, it is convenient to analyze the dynamical behavior of the nucleus
in the body-fixed rotating frame, which can be easily obtained by calculating the principal axes of inertia, i.e., the short, intermediate, and long axes of the nucleus.
At the initial time, the axes of the space-fixed frame are set to be along with the three principal axes of inertia.

If one performs the TDCDFT calculations starting from the lowest energy state for a given spin, e.g., see Fig.~\ref{fig2}(a), the nucleus will rotate around a tilted axis in the short-long plane.
Therefore, in the body-fixed frame, one could see that the angular momentum vector will be always pointing along the tilted axis.
It, however, becomes quite different if the initial state is away from the lowest energy state.

\begin{figure}[!htbp]
  \centering
  \includegraphics[width=0.45\textwidth]{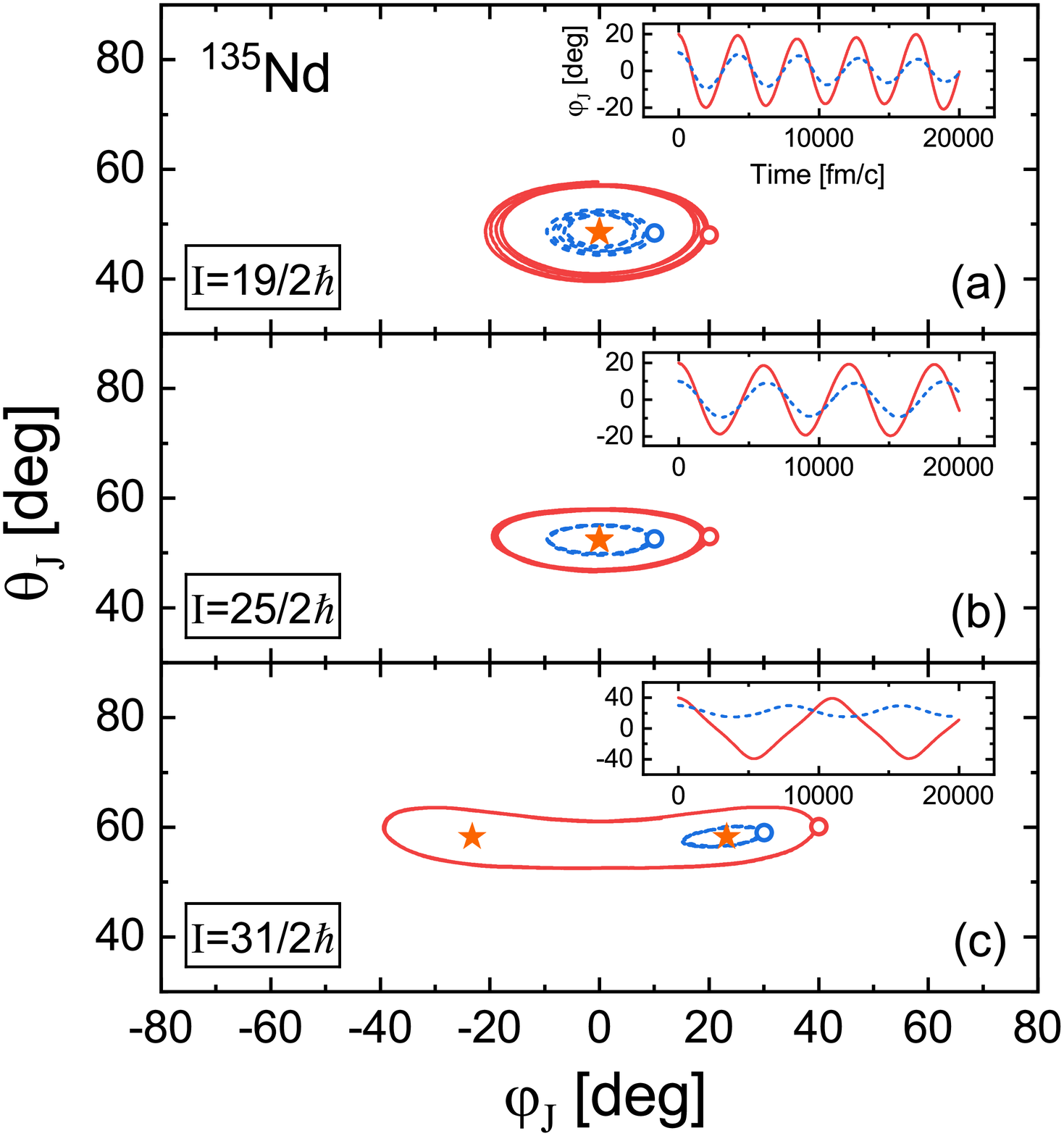}\\
  \caption{(Color online). Trajectories of the tilted angles $(\theta_J, \varphi_J)$ for the total angular momenta in the body-fixed frame by the TDCDFT calculations starting from the initial states with spin $I=19/2\hbar$ (top), $25/2\hbar$ (middle), and $31/2\hbar$ (bottom) obtained from TAC-CDFT calculations with the configuration $\pi[h_{11/2}^1(gd)^1]\otimes\nu h_{11/2}^{-1}$.
  The red solid and blue dash lines correspond to the trajectories starting with different initial states, whose $(\theta_J, \varphi_J)$ values are represented by the open circles.
  The stars denote the location of the minima in the energy surfaces.
  Insets: Time evolutions for the azimuth angle $\varphi_J$.
  }\label{fig3}
\end{figure}

This can be seen in Fig.~\ref{fig3}, where the trajectories of the tilted angles $\theta_J$ and $\varphi_J$ for the angular momenta in the body-fixed frame are depicted for the initial states with $I=19/2\hbar$, $25/2\hbar$, and $31/2\hbar$.
For each spin, two states away from the lowest energy state are employed as the initial states for the TDCDFT calculations.
For $I=19/2\hbar$ and $25/2\hbar$, the trajectories are roughly ellipses centered at the locations of the minima in the corresponding energy surfaces.
This provides clearly a precession picture, i.e., the angular momentum vector itself is rotating around the tilted axis in the short-long plane (see Fig.~\ref{fig1}).
Moreover, the precession is much more extended in the $\varphi_J$ direction than in the $\theta_J$ one.
In the $\varphi_J$ direction, it involves both positive and negative values, which reflects the chiral nature of the precession due to the transitions between the left- and right-handed sectors.
Therefore, we call it ``chiral precession'', which reveals the microscopic dynamics of the chiral excitations; see movies in the Supplemental Material~\cite{[See Supplemental Material for the movies of the chiral precession in both the space-fixed and body-fixed frames from the TDCDFT calculations and for the details on the Fourier analysis to extract the chiral excitation energy] supplement1}.
For a given spin $I=19/2\hbar$ or $25/2\hbar$, although the precession amplitudes starting from the two initial states are quite different, the corresponding precession periods are roughly identical, as shown in the insets of Fig.~\ref{fig3}.
This demonstrates the harmonic nature of the chiral precession, which is consistent with the chiral vibration picture.

Note that the precession mechanism here cannot be deduced only from the energy surfaces shown in Fig.~\ref{fig2}, which provide only the potential part of the chiral motion.
The kinetic part may also play an important role in generating the precession.
Further studies in this direction would be an interesting topic.

For $I=31/2\hbar$, since there are two degenerate energy minima separated by a barrier at $\varphi_J = 0^\circ$, the trajectories strongly depend on the choice of the initial states.
If the initial energy is lower than the barrier, the trajectory is limited in the neighborhood of the minimum with a finite $\varphi_J$ corresponding to either the left- or the right-handed sector.
However, if the initial energy is higher than the barrier, the trajectory could extend across both positive and negative $\varphi_J$ values.
As seen in the inset of Fig.~\ref{fig3}(c), the two precessions are disparate not only in their amplitudes but also in the periods.
This indicates the anharmonic effects of the chiral motion, which are associated with quantum tunneling between left- and right-handed sectors.

To understand this phenomenon further, it would be necessary to resort to the so-called requantization of the TDCDFT~\cite{ring2004nuclear}, e.g., the time-dependent generator coordinate method.
However, to our knowledge, it is hard, if not impossible, to be implemented in a realistic microscopic description of nuclei due to the inaccessible computational requirements.
One of the ways out is probably to derive a collective Hamiltonian with the Gaussian overlap approximation (GOA) in the space of the tilted angles~\cite{Chen20131DCH, Chen20162DCH}.
However, there are still many open questions about the quality of the GOA, the determination of the mass inertia, the influence of the time-reversal symmetry breaking, etc.
The solutions to all these problems are apparently beyond the scope of this work.

\begin{figure}[!htbp]
  \centering
  \includegraphics[width=0.45\textwidth]{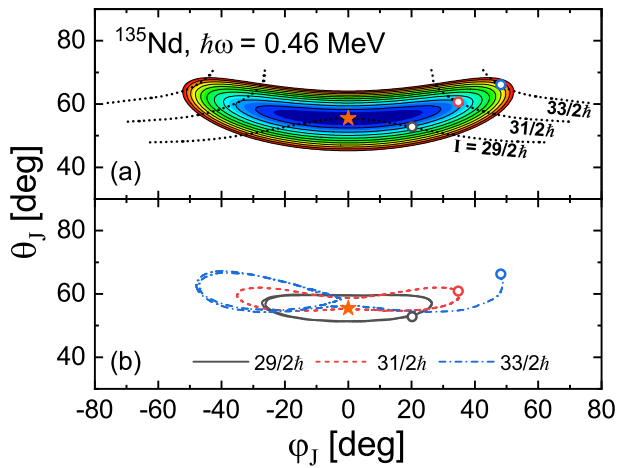}\\
  \caption{(Color online). (a) Same as Fig.~\ref{fig2} but for the total Routhian surface at a given rotational frequency $\hbar\omega=0.46$~MeV, where the spin varies on the surface (dotted lines).
  (b) Same as Fig.~\ref{fig3} but for the initial states with spin $I=29/2\hbar$, $31/2\hbar$, and $33/2\hbar$ obtained from the Routhian surface.
  }\label{fig4}
\end{figure}

It must be emphasized that in many TAC analyses for the nuclear chirality, the calculations are performed for fixed rotational frequencies $\hbar\omega$ but not fixed angular momenta.
This seems to be a natural choice because the angular momentum is anyhow not a good quantum number in the mean-field framework.
Therefore, as depicted in Fig.~\ref{fig4}, we examine the chiral precession trajectories starting from initial states with a given rotational frequency.
In Fig.~\ref{fig4}(a), the total Routhian surface at $\hbar\omega=0.46$~MeV is shown, and there is only one minimum on the surface.
For the TDCDFT calculations, here we take three states on the total Routhian surface as the initial states, which have the same $\hbar\omega$ but different angular momenta.
Since only the average angular momentum is conserved, but not the rotational frequency, during the time evolutions in TDCDFT, the obtained three trajectories are quite different.
Moreover, the trajectories for $I=31/2\hbar$ and $I=33/2\hbar$ indicate that there should be two energy minima, and this is in contradiction with the single minimum in the total Routhian surface.
The trajectory for $I = 29/2\hbar$ is compatible with the single minimum in the Routhian surface, but this is an accident, because the Routhian minimum at $\hbar\omega=0.46$~MeV and the energy minimum at $I=29/2\hbar$ are at the same location in the plane of $\theta_J$ and $\varphi_J$.
This clearly demonstrates the importance of the self-consistent constrained energy surface at a given spin in the TAC-CDFT calculations.

\begin{figure}[!htbp]
  \centering
  \includegraphics[width=0.45\textwidth]{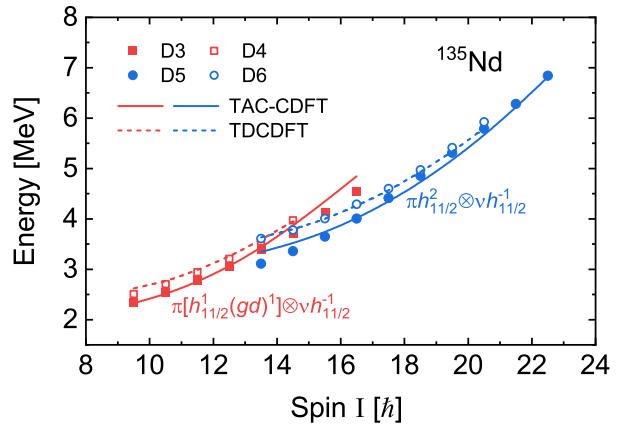}\\
  \caption{(Color online). Calculated excitation energies (solid and dashed lines) for chiral doublet bands in $^{135}$Nd built on the configurations $\pi[h_{11/2}^1(gd)^1]\otimes\nu h_{11/2}^{-1}$ and $\pi h_{11/2}^2\otimes\nu h_{11/2}^{-1}$  in comparison with data (solid and open symbols)~\cite{Lv2019PRCNd135}.
  The solid lines represent the TAC-CDFT results, and the dashed ones represent the results including the chiral excitation energies obtained with the TDCDFT calculations.
  }\label{fig5}
\end{figure}

The energy of the chiral excitation for a given spin can be extracted through the Fourier analysis~\cite{Press1992NumericalRecipes, Smith2002DFT, Reinhar2006boundary} on the time evolution of the chiral precession; see details in the Supplemental Material~\cite{[See Supplemental Material for the movies of the chiral precession in both the space-fixed and body-fixed frames from the TDCDFT calculations and for the details on the Fourier analysis to extract the chiral excitation energy] supplement1}.
This, together with the TAC-CDFT solutions, could provide a microscopic description for the chiral doublet bands.
The Fourier analyses have been done for the states with the spins in the harmonic regime.
In Fig.~\ref{fig5}, the calculated energies for the two pairs of chiral doublet bands in $^{135}$Nd, which are built respectively on the configurations $\pi[h_{11/2}^1(gd)^1]\otimes\nu h_{11/2}^{-1}$ and $\pi h_{11/2}^2\otimes\nu h_{11/2}^{-1}$, are depicted in comparison with the data~\cite{Lv2019PRCNd135}.
It can be seen that the experimental energies are well reproduced.
This provides the first fully microscopic and self-consistent description for the chiral doublet bands in the framework of DFTs.

In summary, for the first time, the dynamics of chiral nuclei has been investigated in the microscopic time-dependent covariant density functional theory with the initial states from the static three-dimensional tilted axis cranking solutions.
The experimental energies of the two pairs of the chiral doublet bands in $^{135}$Nd are well reproduced without any adjustable parameters beyond the well-defined density functional.
Taking the positive-parity bands as an example, a transition from the planar into aplanar rotations has been found from the total energy surfaces with fixed spins.
A novel mechanism, chiral precession, has been revealed from the microscopic dynamics of the total angular momentum in the body-fixed frame.
The obvious planar-aplanar transition corresponds to the increasingly anharmonic effects in the chiral procession.
It is found that, however, such a correspondence is covered up for the total Routhian surfaces at fixed rotational frequencies, since the average angular momentum, rather than the rotational frequency, is conserved during the time evolutions.

\begin{acknowledgments}
  This work was partly supported by the National Key R\&D Program of China (Contracts No. 2018YFA0404400 and 2017YFE0116700), the National Natural Science Foundation of China (Grants No. 12070131001, 11875075, 11935003, 11975031, and 12141501), the China Postdoctoral Science Foundation under Grant No. 2020M670013, and the High-performance Computing Platform of Peking University.
\end{acknowledgments}

\end{document}